\documentclass[twocolumn]{aastex62}
\usepackage{natbib,amsmath}
\usepackage{hyperref} \usepackage{float} 
\usepackage{afterpage}
\usepackage{subfigure}
\usepackage{xspace}

\newcommand{\um}{\textmu m\xspace}
\providecommand{\e}[1]{\ensuremath{\times 10^{#1}}}

\graphicspath{{./}{figures/}}

\begin{document}

\title{GJ 367b is a dark, hot, airless sub-Earth}

\correspondingauthor{Michael Zhang}
\email{mzzhang2014@gmail.com}

\author[0000-0002-0659-1783]{Michael Zhang}
\affil{Department of Astronomy \& Astrophysics, University of Chicago, Chicago, IL 60637}

\author[0000-0003-2215-8485]{Renyu Hu}
\affiliation{Jet Propulsion Laboratory, California Institute of Technology, Pasadena, CA 91109, USA}
\affiliation{Division of Geological and Planetary Sciences, California Institute of Technology}

\author[0000-0001-9164-7966]{Julie Inglis}
\affiliation{Division of Geological and Planetary Sciences, California Institute of Technology}

\author[0000-0002-8958-0683]{Fei Dai}
\affiliation{Division of Geological and Planetary Sciences, California Institute of Technology}
\affiliation{Department of Astronomy, California Institute of Technology, Pasadena, CA 91125, USA}

\author[0000-0003-4733-6532]{Jacob L.\ Bean}
\affil{Department of Astronomy \& Astrophysics, University of Chicago, Chicago, IL 60637}

\author[0000-0002-5375-4725]{Heather A. Knutson}
\affiliation{Division of Geological and Planetary Sciences, California Institute of Technology}

\author[0000-0002-9910-6088]{Kristine Lam}
\affiliation{Institute of Planetary Research, German Aerospace Center (DLR), Rutherfordstrasse 2, D-12489 Berlin, Germany}

\author[0000-0001-9670-961X]{Elisa Goffo}
\affiliation{Dipartimento di Fisica, Universita degli Studi di Torino, via Pietro Giuria 1, I-10125, Torino, Italy}
\affiliation{Thüringer Landessternwarte Tautenburg, Sternwarte 5, D-07778 Tautenburg, Germany}

\author[0000-0001-8627-9628]{Davide Gandolfi}
\affiliation{Dipartimento di Fisica, Universita degli Studi di Torino, via Pietro Giuria 1, I-10125, Torino, Italy}

\begin{abstract}
We present the mid-infrared (5--12 \um) phase curve of GJ 367b observed by the Mid-Infrared Instrument (MIRI) on the James Webb Space Telescope (JWST).  GJ 367b is a hot ($T_{eq}=1370$ K), extremely dense ($10.2 \pm 1.3$ g\,cm$^{-3}$) sub-Earth orbiting a M dwarf on a 0.32 d orbit.  We measure an eclipse depth of $79 \pm 4$ ppm, a nightside planet-to-star flux ratio of $4 \pm 8$ ppm, and a relative phase amplitude of $0.97 \pm 0.10$--all fully consistent with a zero-albedo planet with no heat recirculation.  Such a scenario is also consistent with the phase offset of $11 \pm 5$ degrees East to within 2.2$\sigma$.  The emission spectrum is likewise consistent with a blackbody with no heat redistribution and a low albedo of $A_B \approx 0.1$, with the exception of one anomalous wavelength bin that we attribute to unexplained systematics.  The emission spectrum puts few constraints on the surface composition, but rules out a CO$_2$ atmosphere $\gtrsim$1 bar, an outgassed atmosphere $\gtrsim$10 mbar (under heavily reducing conditions), or an outgassed atmosphere $\gtrsim$0.01 mbar (under heavily oxidizing conditions).  The lack of day-night heat recirculation implies that 1 bar atmospheres are ruled out for a wide range of compositions, while 0.1 bar atmospheres are consistent with the data. Taken together with the fact that most of the dayside should be molten, our JWST observations suggest the planet must have lost the vast majority of its initial inventory of volatiles.
\end{abstract}


\section{Introduction} 
\label{sec:intro}
The question of whether small rocky planets orbiting M dwarfs can host atmospheres is of prime importance for habitability.  Due to the small radius and low luminosity of these stars, such atmospheres would be far easier to study than equivalent atmospheres of planets around Sun-like stars.  However,  it has long been suggested that the high-energy radiation, flares, and long pre-main-sequence of M dwarfs strip planetary atmospheres; the extent to which this happens is a subject of active research (e.g. \citealt{hawley_2014,loyd_2018,gunther_2020,do_Amaral_2022,nakayama_2022}).  Whether small planets can retain atmospheres under such a hostile stellar environment is an open question, because both the processes that create atmospheres (e.g. volatile delivery during planet formation and magma outgassing) and mass loss processes (e.g. photoevaporation and stellar wind erosion)  are poorly understood.  For example, extreme ultraviolet flux is important for driving atmospheric escape, but the EUV flux of even the closest star is uncertain to more than an order of magnitude \citep{france_2022}.  Magnetic fields likely play a crucial role in regulating mass loss, but the strength of exoplanet magnetic fields is unknown, as is whether magnetic fields increase or decrease the mass loss rate \citep{ramstad_2021}.  By observing M dwarf planets and determining which, if any, host atmospheres, we can build up a sample of empirical benchmarks that can be used to calibrate atmospheric mass loss models.

Over the past five years, the Transiting Exoplanet Survey Satellite (TESS) has surveyed nearly the entire sky and discovered a treasure trove of M dwarf rocky planets amenable to study by the recently launched James Webb Space Telescope (JWST).  One such planet is GJ 367b \citep{lam_2021,goffo_2023}, a hot sub-Earth ($R=0.70 R_\Earth$, $M=0.63 M_\Earth$, $T_{eq}=1367$ K) orbiting a M1V star on a 0.32 day orbit.  GJ 367b is by far the most observationally favorable sub-Earth according to the Emission Spectroscopy Metric \citep{kempton_2018}, and even among planets with $R < 1.5 R_\Earth$, it has the second highest ESM \citep{lam_2021}.  As TESS has already surveyed most of the sky multiple times, GJ 367b will likely forever remain the transiting sub-Earth with the highest ESM.

Very little is known about hot sub-Earths.  It is currently unclear whether or not sub-Earth planets form their cores and acquire atmospheres via the same channels as their more massive super-Earth and sub-Neptune cousins, or whether these small planets represent a distinct formation channel with correspondingly different outcomes for their atmospheric and surface properties.  There is some empirical evidence for the latter hypothesis from exoplanet demographics \citep{qian_2021}.  The suggestion by \cite{sinukoff_2013} (inspired by \citealt{cameron_1985,fegley_1987,valencia_2010}) that the hottest sub-Earths may be the remnants of more massive planets whose silicate mantles evaporated away is particularly intriguing because GJ 367b's exceptional density ($10.2 \pm 1.3$ g cm$^{-3}$) is consistent with an iron core comprising $91_{-23}^{+7}$\% of the mass \citep{goffo_2023}.  Another possibility, not exclusive with the first, is that GJ 367b formed like Mercury \citep{benz_1988}: a giant impact stripped the mantle and left the core behind.  This process does not necessarily strip all volatiles, as evidenced by pyroclastic volcanism on Mercury \citep{kerber_2009}.  Finally, \cite{mah_2023} suggested that iron-rich planets can simply be formed from iron-rich materials, as outwardly drifting iron vapour condenses and increases the pebble iron mass fraction.

Thermal emission is a useful tool to probe the potentially exotic surface and atmosphere of GJ 367b and of other rocky planets.  \cite{kreidberg_2019} used Spitzer to observe a phase curve of the super-Earth LHS 3844b, concluding from the undetectable nightside emission and large phase amplitude that the planet is consistent with an airless rock.  On the other hand, the large hot spot offset seen for 55 Cnc e \citep{demory_2016} may suggest atmospheric circulation, although the data analysis has been called into question by \cite{mercier_2022}.  In the JWST era, JWST measured the 15 \um photometric eclipse of TRAPPIST-1b and found it fully consistent with a bare black rock \citep{greene_2023,ih_2023}.  On the other hand, the 15 \um eclipse of TRAPPIST-1c was marginally shallower than the deepest possible, ruling out a 0.1 bar pure CO$_2$ atmosphere or a 10 bar atmosphere with 10 ppm CO$_2$, but not ruling out thinner atmospheres \citep{zieba_2023,lincowski_2023}.

\section{Observations and Reduction}
\label{sec:observations}
Using the Mid-Infrared Instrument \citep{kendrew_2015} on JWST, we monitored GJ 367b for 12.7 h, corresponding to 1.6 planetary orbits (GO 2508, PI: M. Zhang).  These observations were taken in Low Resolution Spectroscopy (LRS) slitless Time Series Observation mode.  Due to the brightness of the star (K=5.8), we used only 5 groups per integration.  This was not enough to avoid saturation in the final group for all pixels, if saturation is defined conservatively at 80\% of full well (the brightest pixel has 58,400 DN, 86\% of full well).  However, we fit the five up-the-ramp reads in such a way that the final group does not bias the slope, and the fourth group does not reach above 52,400 DN (77\% of full well).  The observation consisted of 3 exposures, with a total of 47,919 integrations.

We analyzed the data with the open source package Simple Planetary Atmosphere Reduction Tool for Anyone (SPARTA), first described in \cite{kempton_2023}.  We start from the uncalibrated data and proceed all the way to the final results without using any code from any other pipeline.  We describe the specific pipeline steps, with particular focus on where it differs from \cite{kempton_2023}, in Appendix \ref{sec:appendix_pipeline}.  An independent reduction with Eureka \citep{bell_2023} is also presented in Appendix \ref{sec:appendix_pipeline}; the resulting emission spectrum is fully consistent with our fiducial SPARTA reduction, except for a single wavelength bin.

The unbinned and un-detrended spectroscopic light curves obtained by SPARTA show a variety of systematics (Figure \ref{fig:raw_data}), some of which were also present in the MIRI/LRS phase curves of GJ 1214b \citep{kempton_2023} and/or WASP-43b (ERS Team 2023), others of which make their first unwelcome appearance.  The odd-even effect (manifesting as an alternating bright/dark pattern) and the (quasi-)exponentialy declining ramp fall into the first category, appearing in all MIRI/LRS observations that we are aware of.  The shadowed region between 10.54--11.76 \um has a sharply rising rather than falling ramp, and corresponds to a region of the detector that is unilluminated when the dispersive element is not in the optical path \citep{bell_2024}.  For unknown reasons, the shadowed region is seen in many but not all MIRI datasets.  The odd-even effect of alternating bright/dim columns, especially visible at the beginning, is a feature of every MIRI dataset.  We suspect that it is due to flux redistribution between adjacent pixels, which would also explain why, for the commissioning dataset of a transmission spectrum known to be flat, binning in wavelength improves the accuracy of the result by more than $\sqrt{N}$ (Taylor Bell, private communication).  The notch around 8.1 \um at the very end of the observations has not been seen in previous observations, nor has the sudden 500 ppm drop in flux 1.7 h after the beginning of observations.  The cause of these two anomalies has yet to be determined.

We take a number of measures to mitigate these systematics.  To avoid modelling the flux drop, we cut the first 1.7 h of observations, which also dramatically mitigates the quasi-exponential declining ramp.  To repair the notch, for each spectrum, we replace the fluxes within the notch by the average of the 6 pixels to the left and the 6 pixels to the right.  To mitigate the odd-even effect, we define our wavelength bins to be as wide as possible.  To mitigate the effect of the shadowed region, we only use wavelengths blueward of the region to compute the white light curve.  For the spectroscopic light curves, we define the wavelength bins so that the shadowed region is covered by an integer number of bins; this way, we minimize the number of affected light curves, and avoid the difficulty of modelling ``partially shadowed'' light curves.  Our wavelength bin boundaries are from 5060 \AA{} to 11759 \AA{} inclusive in 609 \AA{} bins (ranging from 9 spectral pixels at the blue end to 33 at the red end), with the reddest two bins (10541--11150 \AA{}, 11150--11759 \AA{}) perfectly spanning the shadowed region.

We use MCMC, as implemented by the \texttt{emcee} package, to fit both the white and spectroscopic light curves.  For both, we use the following systematics model:
\begin{align}
    M_{\rm sys} = F_* (1 + A\exp{(-t/\tau)} + c_y y + c_x x + m(t-\overline{t})),
\end{align}

where A and $\tau$ are the amplitude and timescale of the exponential ramp, y is the position of the trace in the dispersion direction, x is the position of the trace in the spatial direction, and $\overline{t}$ is the average time.  $c_x$, $c_y$, and $m$ are linear decorrelation parameters.

The planet flux model, which we fit simultaneously with the systematics model, is:
\begin{align}
    F_p = E + C_1(\cos(\omega t) - 1) + D_1 \sin(\omega t) + \\
    C_2(\cos(2\omega t) - 1) + D_2 \sin(2\omega t),
    \label{eq_flux}
\end{align}

where E is the eclipse depth, $C_1$, $D_1$, $C_2$, and $D_2$ are phase curve parameters \citep{cowan_agol_2008}, and $\omega = 2\pi/P$.  With this parameterization, the dayside planet-to-star flux ratio is $E$, and the nightside planet-to-star flux ratio is $E - 2C_1$.  We chose a second-order phase curve model because for the mini-Neptune GJ 1214b, we found that theoretical spectroscopic phase curves derived from general circulation models are sometimes poorly fit by a first-order model, but that a second-order model fits all GCMs well \citep{kempton_2023}.  A second-order model fit, by being more flexible, is also more data-driven and therefore more conservative. The dayside planet-to-star flux ratio and nightside planet-to-star flux ratio are both unaffected by the addition of second-order terms: the former because it depends almost entirely on the eclipses, the latter because the nightside ratio depends only on $E$ and $C_1$.

We model the shape of the transit and eclipse using \texttt{batman} \citep{kreidberg_2015}.  The transit and eclipse times are free parameters in the white light fit, but not in the spectroscopic fits.  We fix transit parameters $a/R_*$, P, and inclination to the values found in \cite{goffo_2023}, namely 3.327, 0.3219225 d, and 79.89$^\circ$ respectively; in turn, \cite{goffo_2023} derive their transit parameters from a joint fit with HARPS radial velocities and TESS transit observations.  TESS has observed hundreds of transits, allowing it to measure the transit parameters much more accurately than we can achieve with one JWST/MIRI transit observation.

We analyze the residuals of the white light and spectroscopic phase curves to quantify the correlated noise.  The unbinned white light residuals have a RMS of 227 ppm, 1.12$\times$ the photon noise calculated by our pipeline and 1.16$\times$ the photon noise calculated by PandExo 3 \citep{batalha_2017}.  Binning by a factor of 2048 (0.54 h, close to the eclipse duration of 0.63 h), we obtain a RMS of 14 ppm, 2.7$\times$ the photon noise from the pipeline.  The RMS of the unbinned spectroscopic light curves are a few percent higher than the pipeline photon noise from 5--9 \um, rising to 9\% higher at 10.2 \um and 17\% higher at 11.5 \um.  The pipeline's reported photon noise is 0.95--1.07$\times$ the PandExo prediction at all wavelengths.  After binning by 2048$\times$, the RMS of the spectroscopic light curves is $\sim1.60\times$ the pipeline photon noise at 5 \um, falling to $\sim1.30\times$ at 12 \um.

Part of this excess correlated noise may be due to stellar activity.  Rotational variability is not a concern because the rotational period is 40-50 days, 100x longer than our observations \citep{lam_2021}.  We did not detect any flares in the TESS light curve, to a sensitivity limit of $\sim$700 ppm.  However, JWST is far more sensitive to flares than TESS, and it is possible a 50 ppm mini-flare occurred during our observations (Figure \ref{fig:white_phase_curve}).


\section{Results}
\label{sec:results}
\begin{figure}[htbp]
   \includegraphics[width=0.5\textwidth]{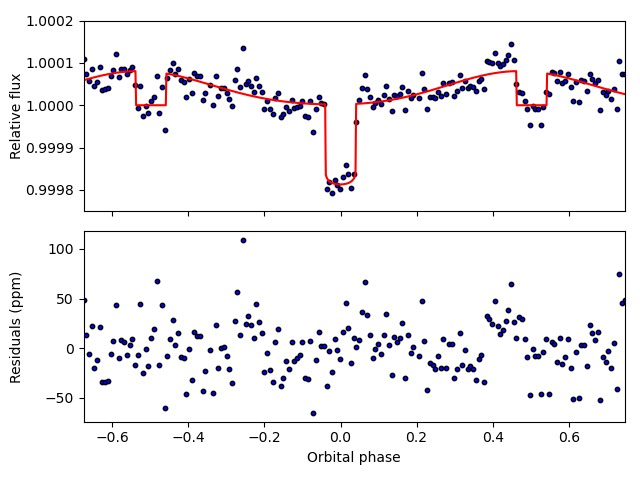}   
    \caption{The white light phase curve (5--10.541 \um), cutting off at the beginning of the shadowed region on the red end. Top: the systematics-corrected data (blue) and the planet flux model fit to the data (red); bottom: the residuals of the fit.  For plotting purposes only, the data are binned into 200 bins, with 206 points in each bin.}
    \label{fig:white_phase_curve}
\end{figure}

\begin{figure*}[ht]
    \subfigure {\includegraphics[width=0.5\textwidth]{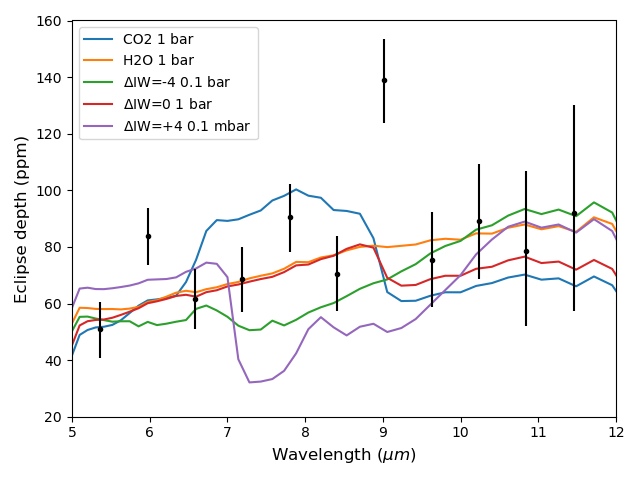}}
    \subfigure {\includegraphics[width=0.5\textwidth]{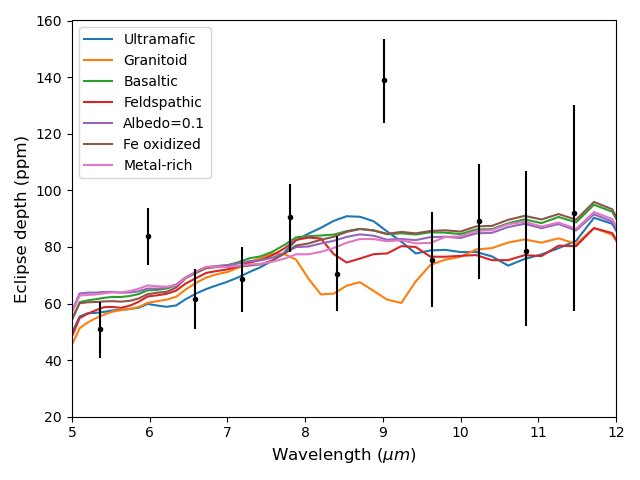}}
    \caption{Observed dayside emission spectrum, compared to selected self-consistent 1D forward models computed by HELIOS 3.1.  The atmosphere models (left) have a metal-rich surface, while the surface models (right) have no atmosphere.  We ignore the outlier 9.0 \um data point when comparing to models (but see Appendix \ref{sec:possible_silicate_feature} for reasons it may be real).}
    \label{fig:eclipse_spectrum}
\end{figure*}

\begin{table}[htbp]
    \caption{Inferred parameters from the white light phase curve.  Upper/lower limits are 95th/5th percentiles.}
    \centering
    \setlength{\tabcolsep}{6pt}
    \begin{tabular}{C C}
        \hline
        \text{Property} & \text{Value}\\
	\hline
        F_{\rm p, day} / F_* \, \text{(ppm)} & 79 \pm 4\\
        F_{\rm p, night} / F_* \, \text{(ppm)} & 4 \pm 8\\
        F_{\rm p, night} / F_{\rm p, day}  & < 0.21\\
        A = (F_{\rm max} - F_{\rm min}) / F_{\rm max} & 0.97 \pm 0.10\\
        \phi \, (^\circ \text{East}) & 10.6 \pm 4.8\\
        $T_e - T_t - P/2$ \, (s) & 52 \pm 13\\
        R_p/R_s & 0.01364 \pm 0.00023\\
        \hline
        T_{\rm day} (K) & 1728 \pm 90\\
        T_{\rm night} (K) & < 847\\
        A_{eff} & 0.01_{-0.23}^{+0.18}\\
        \varepsilon & < 0.16\\        
        f & > 0.602\\
        e\cos{\omega} & 0.0027 \pm 0.0008\\
        \hline
    \end{tabular}
    \label{table:white_phase_curve_params}
\end{table}

Figure \ref{fig:white_phase_curve} shows the systematics-corrected white light phase curve, and Table \ref{table:white_phase_curve_params} shows parameters inferred from it.  We measure a hot day side, a night side consistent with zero emission, and an amplitude consistent with unity.  The phase offset--defined as the phase of the maximum planetary flux minus the phase of the eclipse--is 2.2$\sigma$ eastward of zero.  The phase offset is especially difficult to estimate because inferring it accurately requires accurate modelling of stellar variability, flux variations caused by pointing drift, changes in background flux, and every other systematic effect.  Since there are still unexplained systematics in our data--including the bump before the second eclipse, visible in Figure \ref{fig:white_phase_curve}, that may be a mini-flare--we do not believe a phase offset of zero can be ruled out.  

From the white light phase curve, we find that the eclipse occurs $52 \pm 13$ s later than the transit plus half the orbital period.  Subtracting the light travel time $2a/c$ of 7 s, $45 \pm 13$ s of unexplained delay remains, inconsistent with 0 at the p=0.0035 level (according to the MCMC chains).  This delay can be explained by a slight eccentricity of $e\cos{\omega} = 0.0027 \pm 0.0008$.  A small eccentricity would not be surprising because even though we expect GJ 367b to undergo rapid tidal circularization, it has two super-Earth-mass non-transiting outer companions that can excite eccentricity.  The closer companion has a high eccentricity of $0.23 \pm 0.07$, favorable for exciting GJ 367b's eccentricity (e.g. \citealt{lithwick_2014}).

Following \cite{bolmont_2013}, who in turn use the equations of \cite{leconte_2010}, we calculate the tidal heating due to the planet's eccentricity.  With an Earth-like dissipation factor, tidal heating would equal 5\% of the incident stellar flux at e=0.01, falling to 0.4\% at e=0.0027.  However, GJ 367b could have a dissipation factor orders of magnitude different from Earth's. For example, for 55 Cnc e, \cite{bolmont_2013} consider dissipation factors spanning from $10^{-5}\sigma_p$ to $10^{2}\sigma_p$, where $\sigma_p=1.9\e{-51}$ g$^{-1}$ cm$^{-2}$ s$^{-1}$ is the Earth-like value (and corresponds to $Q_p'=10^4$ for GJ 367b; \citealt{hansen_2010}).  At $10^2\sigma_p$, even an eccentricity of 0.0045 would result in tidal heating that equals the incident stellar flux.  This would cause us the planet to appear much hotter than expected based on simple energy balance assumptions, in contradiction to observations.  However, dissipation factors lower than this extreme value are difficult to rule out with our observations.  The eccentricity damping timescale does not by itself constrain the tidal dissipation factor because of eccentricity pumping from the outer planets.  We ran a coplanar N-body simulation with REBOUND \citep{rein_2012}, initializing planet b with zero eccentricity, and found that b's eccentricity oscillates between $\lesssim$0.0005 and 0.014 with a period of 35,000 yr.  More work will be needed to fully understand the dynamics of this system.

Figure \ref{fig:eclipse_spectrum} shows the emission spectrum, which we obtain by cutting out the two eclipses along with one eclipse duration's worth of baseline on either side, fitting them independently, and averaging the result.  In this way, we are much less affected by non-linearity in the stellar variation (since we fit only a linear slope to the flux), and not affected at all by any instrumental systematics far from the eclipse.  At every wavelength, the spectrum we obtain is within 1$\sigma$ of the one we obtain by fitting the phase curve (Appendix Figure \ref{fig:eclipses_comparisons}, right).  Another advantage of independently analyzing the two eclipses is that we can compare their spectra against each other.  While the eclipse depths are within 1$\sigma$ at most wavelengths, the depths at 6.6 \um are discrepant at 4.3$\sigma$, and the ones at 9.0 \um are discrepant at 2.5$\sigma$.  We have not been able to discover the cause of these discrepancies.

In the combined spectrum, the data point at 9.0 \um is a clear outlier: it is significantly higher than the rest of the spectrum.  No surface or atmospheric model we have tried can produce a feature this sharp.  The non-hydrogen-dominated C/O=0.5 atmosphere model for a 55 Cnc e-like super Earth in \cite{hu_2014} has what appears to be a sharp 40 ppm peak around 9\um, but the sharpness of the peak is a consequence of the higher spectral resolution of the model compared to the data.  Binned to the resolution of the data, the peak becomes a plateau only 10 ppm tall.  Due to the unphysicality of such a sharp feature, and due to the 2.5$\sigma$ inconsistency between the two eclipses at this wavelength, we ignore this data point in our modelling. 
 However, there is an alternate data reduction in which the two eclipses are consistent, suggesting the feature may be real; we discuss this reduction, and the possible interpretation of the feature as silicate emission, in Appendix \ref{sec:possible_silicate_feature}.

Once the 9.0 \um data point is excluded, the emission spectrum is consistent with a featureless blackbody.  Applying nested sampling (using \texttt{dynesty}) to the emission spectrum, we measure the Bond albedo under the assumption of zero heat redistribution, no wavelength dependence in the albedo, and a single average dayside temperature.  We obtain a Bond albedo of $0.11 \pm 0.09$.  Uncertainties on the stellar temperature, on $a/R_*$, and on planet radius are accounted for using Gaussian priors.  The stellar spectrum used to calculate the eclipse depth is obtained from the MIRI observations; the photon noise is negligible, but we assume a 3\% absolute photometric calibration error.  This albedo is consistent within 1.3$\sigma$ of zero, and was calculated under the assumption of no heat redistribution.  In the next section, we relax the assumption, and calculate the albedo and heat redistribution efficiency from the white light phase curve.

In this paper, we focus on analyzing the white-light phase curve and the dayside emission spectrum because we believe these are the most reliable data products.  The transmission spectrum (consistent with being flat), nightside emission spectrum, wavelength-dependent phase curve amplitudes, and wavelength-dependent phase curve offsets are plotted and discussed in Appendix \ref{sec:appendix_spectroscopic_phase_curve}.

\section{Modelling}
\subsection{Albedo and heat recirculation efficiency}
From the white light phase curve, we derive the dayside and nightside brightness temperatures by sampling from the MCMC chain and multiplying the $F_{\rm p,day}/F_*$ and $F_{\rm p,night}/F_*$ by the stellar spectrum observed by MIRI.  We take into account the uncertainty in the stellar effective temperature ($3522 \pm 70$ K) and planet radius ($0.699 \pm 0.024 R_\Earth$), which together contribute a 4\% error.  As shown in Table \ref{table:white_phase_curve_params}, we obtain a nightside temperature consistent with zero and an average dayside brightness temperature of $1709 \pm 92$ K. This is consistent with a zero Bond albedo, zero heat recirculation blackbody, which would have a brightness temperature of $1720 \pm 40$ K.  Since an accurate stellar spectrum is critical to our conclusions, we compared the MIRI-observed stellar spectrum to one computed by interpolation in the SPHINX M-dwarf spectral grid \citep{iyer_2023}, adopting the stellar parameters in \cite{goffo_2023}.  The two spectra were nearly identical, with a typical deviation of 1-3\% and no consistent bias in either direction.

From the dayside and nightside brightness temperatures, we use the toy model of \cite{cowan_2011} to estimate the effective albedo $A_{eff}$ and heat recirculation efficiency $\varepsilon$, obtaining 95th percentile upper limits of $A_{eff}=0.32$ and $\varepsilon=0.17$ respectively.  The parameter $\varepsilon$ ranges from 0 for zero heat redistribution to 1 for full heat redistribution.  The effective albedo is equal to the Bond albedo when the albedo has no wavelength dependence.  However, most realistic surfaces have higher emissivities (and therefore lower $A_{eff}$) in the MIRI bandpass than at either the stellar blackbody peak of 0.8 \um or the planetary thermal blackbody peak of $\sim$2\um.  In fact, \cite{mansfield_2019} found that for all surfaces they considered (metal-rich, Fe-oxidized, basaltic, ultramafic, ice-rich, feldspathic, granitoid, and clay), $A_{eff} < A_B$ by up to 0.3.  The low albedos of most surfaces at MIRI wavelengths mean that the high dayside brightness temperature across the MIRI bandpass is consistent with a wide range of possible surface types.

In addition to calculating the heat redistribution efficiency $\varepsilon$, we calculate the $f$ factor, defined as \citep{burrows_2014}:

\begin{align}
    T_{\rm d,b} = T_s\sqrt{\frac{R_s}{a}} (1-A_B)^{1/4} f^{1/4},
\end{align}

where $T_{\rm d,b}$ is the dayside brightness temperature.  Comparing to \cite{cowan_2011}, we see that $f = \frac{2}{3} - \frac{5}{12}\varepsilon$.  That is, if there is no heat redistribution, $\varepsilon=0$ and $f=2/3$; if there is perfect heat redistribution, $\varepsilon=1$ and $f=1/4$.

\subsection{Self-consistent 1D forward models for the dayside emission spectrum}
\label{sec:forwardmodels}

\begin{table}[htbp]
    \caption{Compatibility of selected HELIOS models of the dayside to the observed dayside spectrum.}
    \centering
    \setlength{\tabcolsep}{6pt}
    \begin{tabular}{c C C C c}
        \hline
        Model & P_{\rm atm} \text{(mbar)} & \chi^2 & \Delta \ln(Z) & Consistent? \\
	\hline
        CO$_2$ & 1000 & 16.9 & -4.1 & N\\
        CO$_2$ & 100 & 14.7 & -2.8 & Y\\
        H$_2$O & 1000 & 8.7 & -0.1 & Y\\        
        $\Delta$IW=-4 & 100 & 21.3 & -4.4 & N\\
        $\Delta$IW=-4 & 10 & 19.7 & -4.2 & N\\
        $\Delta$IW=-4 & 1 & 14.8 & -2.9 & Y\\      
        $\Delta$IW=0 & 1000 & 9.7 & 0.0 & Y\\        
        $\Delta$IW=+4 & 100 & 47.9 & -13.5 & N\\
        $\Delta$IW=+4 & 0.1 & 30.5 & -10.1 & N\\
        $\Delta$IW=+4 & 0.01 & 18.0 & -4.6 & N\\
        $\Delta$IW=+4 & 0.002 & 11.5 & -1.6 & Y\\
        \hline
        Ultramafic & 0 & 9.5 & -0.6 & Y\\
        Granitoid & 0 & 8.4 & 0.1 & Y\\
        Basaltic & 0 & 8.2 & 0.0 & Y\\
        Feldspathic & 0 & 7.3 & 0.5 & Y\\
        Albedo=0 & 0 & 12.6 & -1.0 & Y\\
        Albedo=0.1 & 0 & 8.4 & -0.2 & Y\\
        Iron oxidized & 0 & 8.7 & -0.3 & Y\\
        Metal-rich & 0 & 8.0 & 0 & Y\\
        \hline
    \end{tabular}
    \label{table:helios_models}
\end{table}

We use HELIOS 3.1 \citep{malik_2017,malik_2019a,malik_2019b,whittaker_2022} to self-consistently model the day side of the planet under a variety of assumptions.  We fix the heat redistribution factor $f$ to 2/3, representing zero heat redistribution, in order to explore  the constraints imposed by the (lack of) spectral features.  First, we run airless models with a variety of surfaces using wavelength-dependent albedos from \cite{mansfield_2019} (excluding clay and water ice surfaces because they are implausible for our planet), who in turn take their data from \cite{hu_2012}. In the solar system, basaltic crusts are common, feldspathic crust is found in the lunar highlands, iron oxides are found in abundance on Mars, and granitoid crust, although common on Earth, is less likely on GJ 367b because it requires water to form \citep{campbell_1983}.  Ultramafic rocks were common on the primary Earth and Mars, while metal-rich crusts are unknown in the solar system.

To compare HELIOS models to data, we use nested sampling (implemented by \texttt{dynesty}) with a single parameter, a scaling factor that divides the eclipse depths from HELIOS before comparing with the data and calculating the log likelihood.  The scaling factor accounts for uncertainties in the stellar flux measured by MIRI, in the stellar flux received by the planet, and in the area of the planet's radiating surface.  We give the scaling factor a Gaussian prior with a mean of 1 and standard deviation of 0.077.  The 7.7\% uncertainty is the quadrature sum of a 1.6\% uncertainty on the semimajor axis (which in turn is from a 2.4\% uncertainty on the stellar mass, propagated through Kepler's third law), a 6.9\% uncertainty on the planet cross-sectional area (mostly from the stellar radius uncertainty), and an assumed 3\% flux calibration error.  For each model, we calculate the log of the Bayes factor compared to the fiducial model, an airless world with a metal-rich surface.  Models with $\Delta \ln(Z) < -3$ (corresponding to a Bayes factor of 0.05) are considered inconsistent with the data.
 
As shown in Table \ref{table:helios_models}, all surface compositions we tried are consistent with the data, largely because all have low albedos (high emissivities) in the MIRI bandpass.  Airless worlds with a wavelength-independent albedo of 0 or 0.1 and a corresponding blackbody emission spectrum are also consistent with the data.  Our observations therefore do not constrain the surface composition.  Many complications make our observations even less constraining on the surface composition than the $\chi^2$ numbers imply \citep{mansfield_2019}.  The airless rocky solar system bodies are generally dark, with Bond albedos of 0.07 for Mercury, 0.11 for the Moon, and 0.03 for Ceres.  Graphite contaminants, space weathering, and surface roughness could all decrease the albedo below the already low values measured in the lab (e.g. \citealt{brunetto_2015,dumusque_2019,mansfield_2019}).  On the other hand, if the surface were smoother and had fewer non-metallic elements, it could have a higher albedo than the compositions we assumed.

One major complication which we ignore is that GJ 367b is hot enough to melt most plausible crusts over at least some regions of the day side.  In the case of zero Bond albedo and no heat recirculation, the temperature at the substellar point would be 1930 K; by comparison, iron melts at 1811 K, while silicate rocks start melting at 850 K, and all silicate rocks are molten by $\sim$1500 K \citep{lutgens_2015}.  88\% of the dayside planetary disk has a temperature higher than 1500 K.  Molten lava has different spectral properties from its solid counterpart.  \cite{essack_2020} found that lava worlds should have low albedos ($A_g \lesssim 0.1$), in agreement with our observations.

In addition to computing airless models, we also use HELIOS to compute a variety of atmospheric models of different compositions and thicknesses.  All models adopt the ``metal-rich surface''.  We test a pure CO$_2$ composition, a pure H$_2$O composition, and three possible atmospheres in equilibrium with magma having different oxidation states: $\Delta IW$=-4, 0, and 4, corresponding to highly reduced, Earth-like, and highly oxidized magma respectively.  The outgassed atmospheric compositions were taken from \cite{gaillard_2022} (their Figure 3), and represent the equilibrium state between a magma ocean at 1500$^\circ$C and an atmosphere.  The reduced atmosphere is 56\% H$_2$, 41\% CO, 1.4\% CH$_4$, 0.89\% N$_2$, 0.52\% H$_2$O, 0.1\% CO$_2$, and 0.016\% H$_2$S; the $\Delta IW=0$ atmosphere is 74\% CO, 19\% CO$_2$, 3.3\% N$_2$, 1.9\% H$_2$, 1.8\% H$_2$O, and 0.53\% H$_2$S; and the oxidized atmosphere is 58\% CO$_2$, 35\% SO$_2$, 2.6\% N$_2$, 2.3\% CO, 0.9\% H$_2$O, and 90 ppm H$_2$.  We note that \cite{gaillard_2022} finds an atmosphere of nearly 100 bar for a volatile content similar to that of bulk silicate Earth, while we consider atmospheres orders of magnitude thinner.  We thus implicitly assume the composition does not change during atmospheric escape.  Due to the lack of optical absorbers in our models, none of the HELIOS models have a thermal inversion.

Table \ref{table:helios_models} shows that the emission spectrum places stringent constraints on the maximum atmospheric pressure for some compositions, while placing no constraints for others.  A thick, 1 bar CO$_2$ atmosphere is marginally disfavored by the data, but thinner atmospheres are fully consistent with the emission spectrum.  H$_2$O atmospheres of all thicknesses are also fully consistent.  The emission spectrum does, however, constrain outgassed reduced atmospheres to below 10 mbar, and outgassed oxidizing atmospheres to below 0.01 mbar.  The stringent constraints come from the strong spectral absorption features that these atmospheres produce: largely from CH$_4$ and H$_2$O for the reduced atmosphere, and from CO$_2$ and SO$_2$ for the oxidized atmosphere.

As always, there are many potential caveats to 1D models.  Winds, day-night heat transport, and photochemistry could all alter the temperature-pressure profile.  Clouds can also alter the temperature-pressure profile in addition to attenuating spectral features, potentially making a much larger range of compositions consistent with a blackbody spectrum.  A range of cloud species have condensation temperatures around the dayside temperature of GJ 367b, including iron, silicates, titanium oxides, and calcium titanates (c.f.. \citealt{lodders_2002,wakeford_2017}).  It is possible that some of these clouds could be ruled out based on albedo.  We leave the modelling of these complicating features to future work.

\subsection{From lack of heat recirculation to upper limits on atmospheric thickness}
The dayside emission spectrum does not allow us to rule out thick H$_2$O atmospheres or thick outgassed atmospheres with Earth-like oxidation states, and is barely inconsistent with a 1 bar CO$_2$ atmosphere.  These atmospheres may not have strong spectral features, but they should still transport heat from day to night, and sufficiently thick atmospheres should have heat redistribution efficiencies inconsistent with the white light phase curve observations.  We therefore use the semi-empirical relation found by \cite{koll_2022} to convert from the lack of observed day-night heat transport to upper limits on surface atmospheric pressure.

\cite{koll_2022} modelled global circulation in tidally locked rocky planet atmospheres as a heat engine.  He compared the results to general circulation models (GCMs) to obtain an equation that relates the surface pressure $p_s$ and surface longwave optical depth $\tau_{LW}$ to the heat redistribution factor $f$ (their Equation 10).  We rewrite the equation more simply in terms of $\varepsilon$:

\begin{align}
    \varepsilon = \frac{1}{1 + k/a},
\end{align}

where:
\begin{align}
    a = \tau_{LW}^{1/3} \Big(\frac{p_s}{1 bar} \Big)^{2/3} \Big(\frac{T_{eq}}{600 K}\Big)^{-4/3}
\end{align}

The constraint we obtain from the white light curve, $\varepsilon < 0.17$, implies $a > 0.2k$.  The $k$ parameter, which captures all planetary parameters other than optical thickness, surface pressure, and equilibrium temperature, depends weakly on a planet's exact properties.  From his GCMs, \cite{koll_2022} calculated $k=1.2$ for TRAPPIST-1b, $k=1.9$ for GJ 1132b, and $k=2.3$ for LHS 3844b.  LHS 3844b is the hottest and fastest rotating of the three planets ($T_{eq}=805$ K, $P=0.46$ d), although it is still far colder than GJ 367b ($T_{eq}=1367$ K).  Since $k \propto M/R$, we scale from the value for LHS 3844b to obtain $k=1.2$.  Adopting this value implies that $\tau_{LW} p_s / (1 bar) < 0.4$.  We use the wavelength-dependent optical depths reported by our HELIOS models (see previous subsection) to calculate $\tau_{LW}$ for atmospheres of different compositions and pressures, following Equation 13 from \cite{koll_2022}.  No wavelength cutoff was imposed, but since Equation 13 weights by the Planck function, opacities too far from the blackbody peak of 1.7 \um are effectively disregarded.  We find that $\tau_{LW} p_s / (1 bar) < 0.4$ places the following upper limits on $p_s$: 0.6 bar for a CO$_2$ atmosphere, 0.3 bar for a H$_2$O atmosphere, 0.4 bar for a ``neutral'' outgassed atmosphere, 0.3 bar for a reduced outgassed atmosphere, 0.4 bar for an oxidized outgassed atmosphere.  For a wide range of atmospheric compositions, therefore, the non-detection of heat recirculation in the white light phase curve constrains the surface atmospheric pressure to $\lesssim$0.5 bar.

\subsection{Volatile budget of the planet}
As mentioned in Section~\ref{sec:forwardmodels}, GJ~367~b should have a largely molten dayside. As a result, any volatiles (C, H, N, S) in the silicate mantle of the planet (at least the molten part) should be partitioned into the atmosphere. In other words, the planet must have an atmosphere, unless its bulk is free of any volatiles. For example, using the magma ocean -- atmosphere equilibrium model of \citet{gaillard_2022}, we find that the surface pressure would be 30 -- 100 bars for a volatile content similar to that of bulk silicate Earth and a magma ocean that is 10\% the mass of the planet with an oxygen fugacity $-6<\Delta{\rm IW}<4$. Further reducing the mass of the magma ocean to only 1\% of the planet's mass would still yield an atmosphere of 6 -- 12 bars. It is thus clear that the observed lack of atmosphere indicates that the planet has considerably less volatiles as a whole compared to Earth.

Could atmospheric escape cause this planet-scale volatile depletion, or did the planet have to be formed dry? We estimate the total mass loss with an energy-limited escape rate, similar to the procedure in \citet{hu_2023}. With an escape efficiency of 10\% and an X-ray and EUV luminosity of $10^{28}$ erg s$^{-1}$ (Youngblood, private communication; HST GO 16701 \citealt{youngblood_2021}), 34\% of the planet's mass would escape within the estimated age of 5 billion years \citep{goffo_2023}. Even a lower efficiency of 1\% would be able to remove a volatile content corresponding to $\sim4\%$ of the planet's mass.  Taking into account the decrease over time of stellar XUV would further increase the cumulative mass loss.  The lack of an atmosphere can thus be explained as the removal of initial volatiles by intense stellar irradiation, and is fully consistent with the assessment that the planet is far above the ``cosmic shoreline'' \citep{zahnle2017cosmic}.

\section{Planet in context}
\begin{figure*}[ht]
    \subfigure {\includegraphics[width=0.5\textwidth]{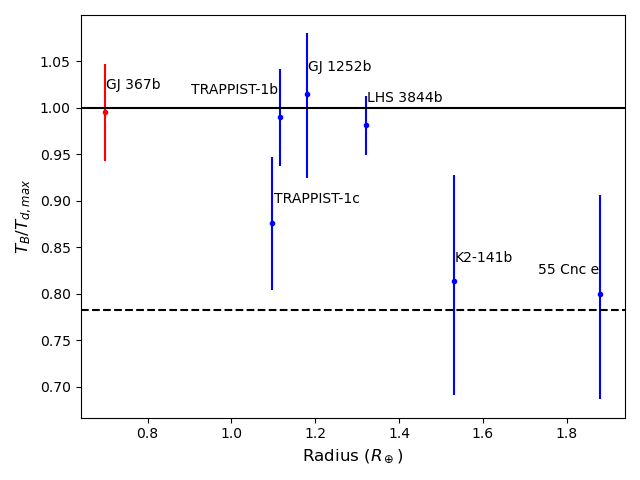}}
    \subfigure {\includegraphics[width=0.5\textwidth]{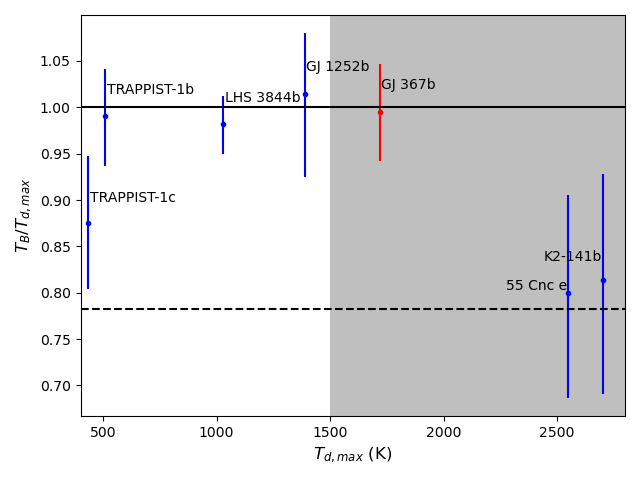}}
    \caption{GJ 367b in the context of other rocky planets with thermal emission measurements.  $T_{rm d,max}$ is the maximum possible dayside temperature, assuming a zero albedo, zero heat redistribution planet.  The dashed horizontal line is at 0.78, the expected ratio for a zero albedo, maximum heat redistribution planet.  The brightness temperature values come from Spitzer IRAC 4.5 \um for GJ 1252b, LHS 3844b, K2-141b, and 55 Cnc e \citep{crossfield_2022}, and from MIRI F1500W photometry for TRAPPIST-1b \citep{greene_2023} and TRAPPIST-1c \citep{zieba_2023}.  On the right, the grey region indicates temperatures high enough to melt all silicate rocks.}
    \label{fig:context}
\end{figure*}

GJ 367b is the first sub-Earth with measured thermal emission, and one of a small but growing database of rocky planets with such measurements.  Figure \ref{fig:context}, an update to Figure 8 of \cite{crossfield_2022}, plots it in the context of these other planets.  GJ 367b extends the very tentative trend of smaller planets having high dayside brightness temperatures relative to the maximum possible.  Such a trend would not be surprising because larger planets generally have higher escape velocities, allowing them to retain an atmosphere more easily.  GJ 367b also begins to fill in the large gap in the $T_B / T_{d,max}$ vs. $T_{d,max}$ plot between 1390 K and 2550 K.

If the very tentative trend of more irradiated planets having lower $T_B / T_{d,max}$ is real, it could indicate that the most irradiated planets have mineral atmospheres--the equilibrium pressure of a mineral atmosphere on a magma world is exponentially sensitive to temperature.  However, it is also worth noting that the two planets with lower $T_B/T_{d,max}$ both orbit hotter stars (G for 55 Cnc e, K for K2-141b), while all other planets orbit M dwarfs. 

The apparent trends in Figure \ref{fig:context} can also easily be coincidence, as there are too few data points to draw reliable conclusions. 
 Much more data are needed to untangle the impact of radius, equilibrium temperature, and stellar type on the existence of an atmosphere.  Fortunately, JWST will vastly expand the number of rocky planets with emission detections.   
 The smaller error bars, larger sample size, and spectroscopic information unavailable with warm Spitzer will provide a much better picture of the surfaces and atmospheres of small rocky planets.  Among other planets, JWST will observe 55 Cnc e (GO 1952, 2084), LHS 3844b (GO 1846, 4008), K2-141b (GO 2347), LTT 1445Ab (GO 2708), GJ 1132b (GTO 1274) GL 486b (GO 1743), and TOI 2445Ab (GO 3784).  Most of these observations will be emission spectra (e.g. 55 Cnc e, LHS 3844b LTT 1445Ab, GJ 1132b, GL 486b), but some will be spectroscopic phase curves (LHS 3844b, K2-141b, TOI 2445Ab).  Most observations will be taken with MIRI/LRS, but some will be taken with other instruments (e.g. NIRCAM/F444W for half of one 55 Cnc e program, NIRSpec/PRISM for TOI 2445Ab).  
 
 Different wavelengths and observing methodologies provide different constraints on surfaces and atmospheres.  Emission spectroscopy can reveal if the dayside brightness temperature is the maximum possible, which would suggest a bare black rock.  If it is lower than the maximum, phase curves are necessary to tell the difference between high Bond albedo and high heat recirculation.  MIRI/LRS is a popular instrument-mode combination for the thermal emission of rocky planets because it maximizes S/N: eclipse depths generally rise toward redder wavelengths, but begin to plateau at LRS wavelengths while the stellar spectrum falls.  On the other hand, most plausible surface compositions have low (and therefore similar) albedos in the MIRI/LRS bandpass, suggesting that bluer wavelengths may be better for distinguishing between different surface compositions.

\section{Conclusion}
\label{sec:conclusion}
GJ 367b is the first sub-Earth with thermal emission observations.  These observations reveal a planet with no detectable atmosphere, no heat redistribution, and a dark surface in the MIRI bandpass ($A_B \approx 0.1$) with a blackbody emission spectrum.  The lack of heat redistribution rules out $\gtrsim$1 bar atmospheres for a wide range of compositions, while the emission spectrum rules out even thinner atmospheres for some compositions.  Given that the planet is far above the ``cosmic shoreline'', the lack of an atmosphere is not surprising, although it is not the best possible news for the prospect of measuring the atmospheres of M dwarf rocky planets.  We encourage JWST observations of planets closer to or below the cosmic shoreline to understand which, if any, rocky planets orbiting M dwarfs have atmospheres.

\textit{Software:}  \texttt{numpy \citep{van_der_walt_2011}, scipy \citep{virtanen_2020}, matplotlib \citep{hunter_2007}, emcee \citep{foreman-mackey_2013}, \citep{speagle_2019}, Eureka! \citep{bell_2022}, SPARTA \citep{kempton_2023}, REBOUND \citep{rein_2012}, HELIOS \citep{whittaker_2022}}

\section{Acknowledgments}
MZ acknowledges support from the 51 Pegasi b Fellowship funded by the Heising-Simons Foundation.  Support for this work was provided by NASA through grant number JWST GO 2508 from STScI. 

E.G. acknowledges the generous support from Deutsche Forschungsgemeinschaft (DFG) of the grant HA3279/14-1.

Some/all of the data presented in this article were obtained from the Mikulski Archive for Space Telescopes (MAST) at the Space Telescope Science Institute. The specific observations analyzed can be accessed via \dataset[DOI: 10.17909/zrda-a787]{https://doi.org/10.17909/zrda-a787}.

We thank our colleagues for fruitful discussions, particularly Austin Stover and the ERS team.


\appendix

\section{Pipeline steps}

\label{sec:appendix_pipeline}
\begin{figure*}[ht]
    \subfigure {\includegraphics[width=0.5\textwidth]{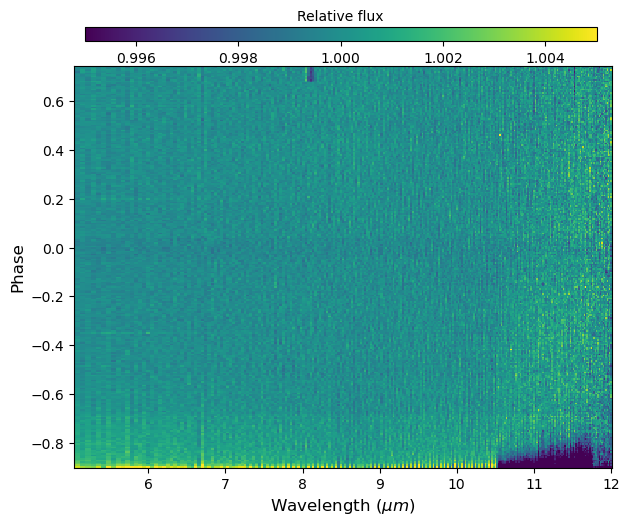}}
    \subfigure {\includegraphics[width=0.5\textwidth]{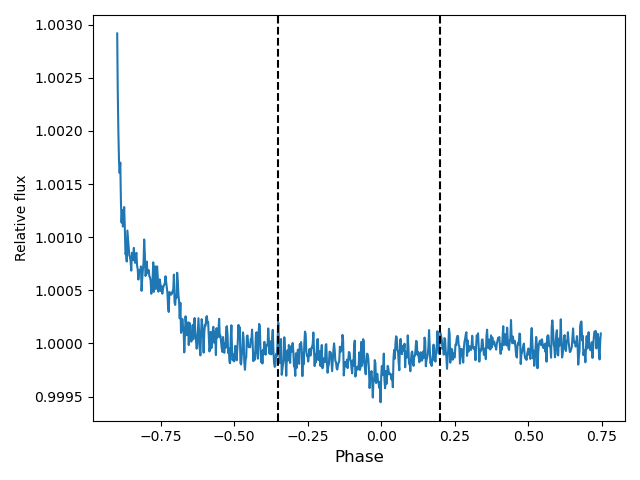}}
    \caption{The data before removal of systematics.  Left: spectroscopic light curves for each wavelength.  The dark band at phase 0 is the transit, and it is the only evident astrophysical feature in this plot.  Note the shadowed region ($\lambda$=10.54--11.76 \um), the notch between $\lambda$=8.04-8.14 \um and phases 0.68--0.74, the sudden flux drop at phase -0.685, and the odd-even effect.  Right: the white light curve.  Note the strong initial ramp and the sudden flux drop at phase -0.685.  The vertical lines mark the boundaries between exposures.}
    \label{fig:raw_data}
\end{figure*}

A summary of the SPARTA pipeline reduction, focusing on where it differs from \cite{kempton_2023}, is as follows:

\begin{enumerate}
    \item calibrate.py performs nonlinearity correction, dark subtraction, multiplication by the gain (assumed to be 3.1 electrons/DN; private comm., Sarah Kendrew), up-the-ramp fitting with outlier rejection (two rounds), and flat fielding, in that order.  We discard the last group in the first round of fitting, calculate the median residuals across all integrations, and add the residuals to the original data (including the last group), thus linearizing it and working around imperfect nonlinearity correction.  We then do a second round of fitting, including all groups.
    \item remove\_bkd.py calculates and removes the background for each row of each integration.  The background is defined as the average of columns 10:25 and -25:-10 (Python indexing convention).
    \item get\_positions.py calculates the position offset of the trace in each integration
    \item get\_med\_image.py calculates the median image across all integrations
    \item optimal\_extract.py uses the median image and the position offset to calculate the profile, and uses the profile to perform optimal extraction.  We use a window with half-width of 5 pixels, and reject pixels more than 5$\sigma$ away from the model image as outliers.
    \item gather\_and\_filter.py gathers the positions and extracted spectra into one file.  Integrations corresponding to 4$\sigma$ outliers in the white light curve are rejected, while 4$\sigma$ outliers in the unbinned spectroscopic light curves are repaired.
\end{enumerate}

\begin{table}[ht]
    \caption{Systematics and mitigation strategies}
    \centering
    \setlength{\tabcolsep}{6pt}
    \begin{tabular}{c c c}
    \hline
    Systematic & Mitigation & Modelling\\
    \hline
    Sudden unexplained flux drop & Trim everything before drop & N/A\\
    Ramp & Trim 100 min at beginning & fit exponential\\
    Non-linearity (RSCD, brighter-fatter) & N/A & Two-step up-the-ramp fitting \\
    ``Notch'' & Replace by average of neighboring wavelengths & N/A\\
    Trace movement & N/A & Detrend with x and y\\
    Subpixel sensitivity variations & N/A & Detrend with x and y\\
    Odd-even effect & Average adjacent spectral pixels (white LC) & N/A\\
    Correlated noise in wavelength & Use big wavelength bins & N/A\\
    Shadowed region & Define wavelength bins that perfectly divide region & N/A\\
    Stellar variability & Cut out eclipses to compute eclipse spectrum & Fit linear function of time\\
    Mini-ramps after exposure gaps & N/A & Mask 170 s\\
    \hline
    \end{tabular}
    \label{table:systematics}
\end{table}

The final output of these steps is shown in Figure \ref{fig:raw_data}.  Not visible in the binned light curve is the fact that the observations were divided into three exposures, with a 40 s gap in between adjacent exposures.  A 1000 ppm mini-ramp can be seen at the start of the second and third exposures, which fades to imperceptibility within 100 seconds.  We therefore mask the first 170 s of data at the start of the second and third exposures.  We tried not masking the mini-ramps, and found a negligible impact on all parameters.

For reader convenience, we summarize the systematics we attempt to mitigate or model in Table \ref{table:systematics}.

\section{The emission spectrum}

\begin{table}[htbp]
    \caption{The fiducial emission spectrum}
    \centering
    \setlength{\tabcolsep}{6pt}
    \begin{tabular}{c c c c}
        \hline
        $\lambda_{min}$ (\um) & $\lambda_{max}$ (\um) & Depth (ppm) & Error (ppm) \\
        \hline
    5.060 &  5.669 & 51 & 10 \\
    5.669 &  6.278 & 84 & 10 \\
    6.278 &  6.887 & 62 & 11 \\
    6.887 &  7.496 & 69 & 12 \\
    7.496 &  8.105 & 90 & 12 \\
    8.105 &  8.714 & 71 & 13\\ 
    8.714 &  9.323 & 139 & 15\\ 
    9.323 &  9.932 & 76 & 17 \\ 
    9.932 &  10.541 & 89 & 20 \\ 
    10.541 & 11.15 & 79 & 28 \\ 
    11.15 &  11.759 & 92 & 38 \\ 
        \hline
    \end{tabular}
    \label{table:emission_spectrum}
\end{table}

\begin{figure*}[ht]
    \subfigure {\includegraphics[width=0.5\textwidth]{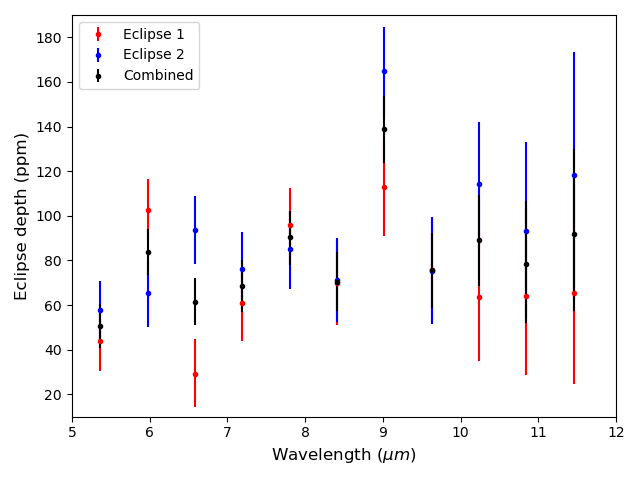}}
    \subfigure {\includegraphics[width=0.5\textwidth]{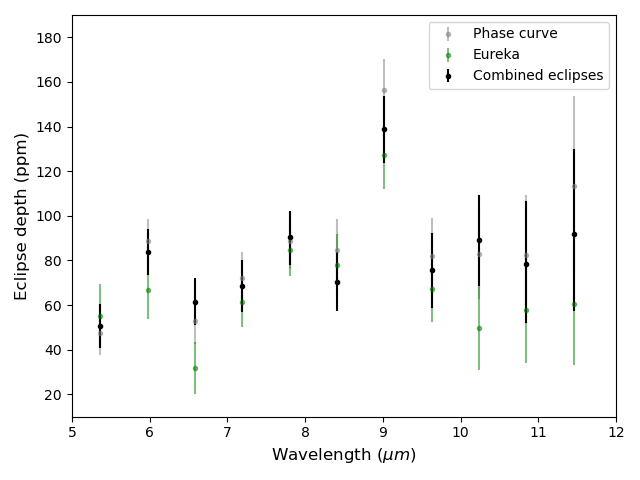}}
    \caption{Left: eclipse depths inferred from the two individual eclipses, together with their average, which we adopt as the fiducial emission spectrum.  Right: the average spectrum of the two eclipses, compared to the spectrum inferred by fitting the entire phase curve simultaneously.  The two are consistent to 1$\sigma$ at all wavelengths.}
    \label{fig:eclipses_comparisons}
\end{figure*}

The fiducial emission spectrum is shown in Figure \ref{fig:eclipse_spectrum} and tabulated in Table \ref{table:emission_spectrum}.  Figure \ref{fig:eclipses_comparisons} shows variants of the emission spectrum.  Of particular note is the spectrum obtained by Eureka \citep{bell_2022}, a completely independent pipeline.  The Eureka reduction uses version 0.9 of the Eureka pipeline, 1.8.2 of the JWST pipeline package, and version 11.17.2 of the CRDS pipeline. The Eureka reduction follows the same steps as with other data sets, with some changes. Eureka utilizes Stage 1 and Stage 2 from the JWST pipeline. The jump step in Stage 1 was skipped as it can introduce noise for small group numbers, as well as the photon flux calibration step in Stage 2, to improve estimation of the flux errors. In Eureka Stage 3, the 2D spectra are rotated counter-clockwise so that the dispersion axis is aligned with the x axis, allowing for reuse of Eureka's functions. A window with a width of 8 pixels centered on the trace was used for spectral extraction. The PSF position and width was recorded to detrend against. Sigma clipping (5$\sigma$) was used to mask out the detector cruciform artifact to avoid biasing the background subtraction. Background subtraction was performed by masking a window of 20 pixels centered on the trace and then fitting a column-wise linear trend to the background to remove the 1/f correlated noise along the cross dispersion axis of the detector, as well as other time correlated noise sources. Eureka Stage 5 was used to fit the phase curve with the same planet flux and systematics models as above
The Eureka reduction was run by JI, independently of the SPARTA reduction run by MZ.  As can be seen in Figure \ref{fig:eclipses_comparisons}, the Eureka emission spectrum matches the fiducial spectrum to 1$\sigma$ at all wavelengths except 6.6 \um, at which they match to within 2$\sigma$.

\section{White phase curve posterior distributions}
\begin{figure*}[ht]
    \centering
    \includegraphics[width=0.5\textwidth]{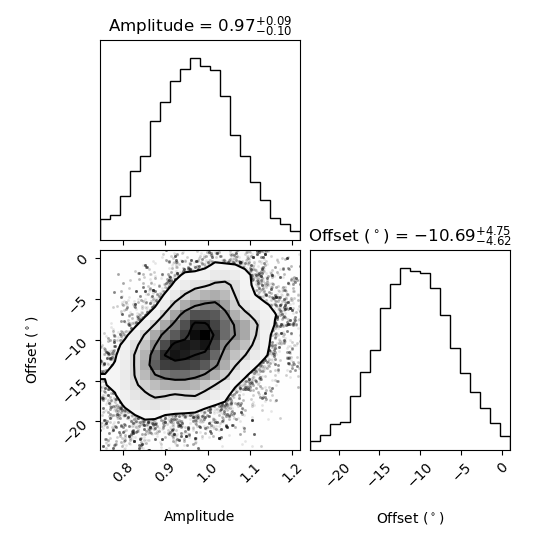}
    \caption{2D posterior distribution (``corner plot'') of the amplitude and phase offset (negative means east) from the white phase curve MCMC fit.}
    \label{fig:amplitude_offset_corner}
\end{figure*}

\begin{figure*}[ht]
    \centering
    \includegraphics[width=\textwidth]{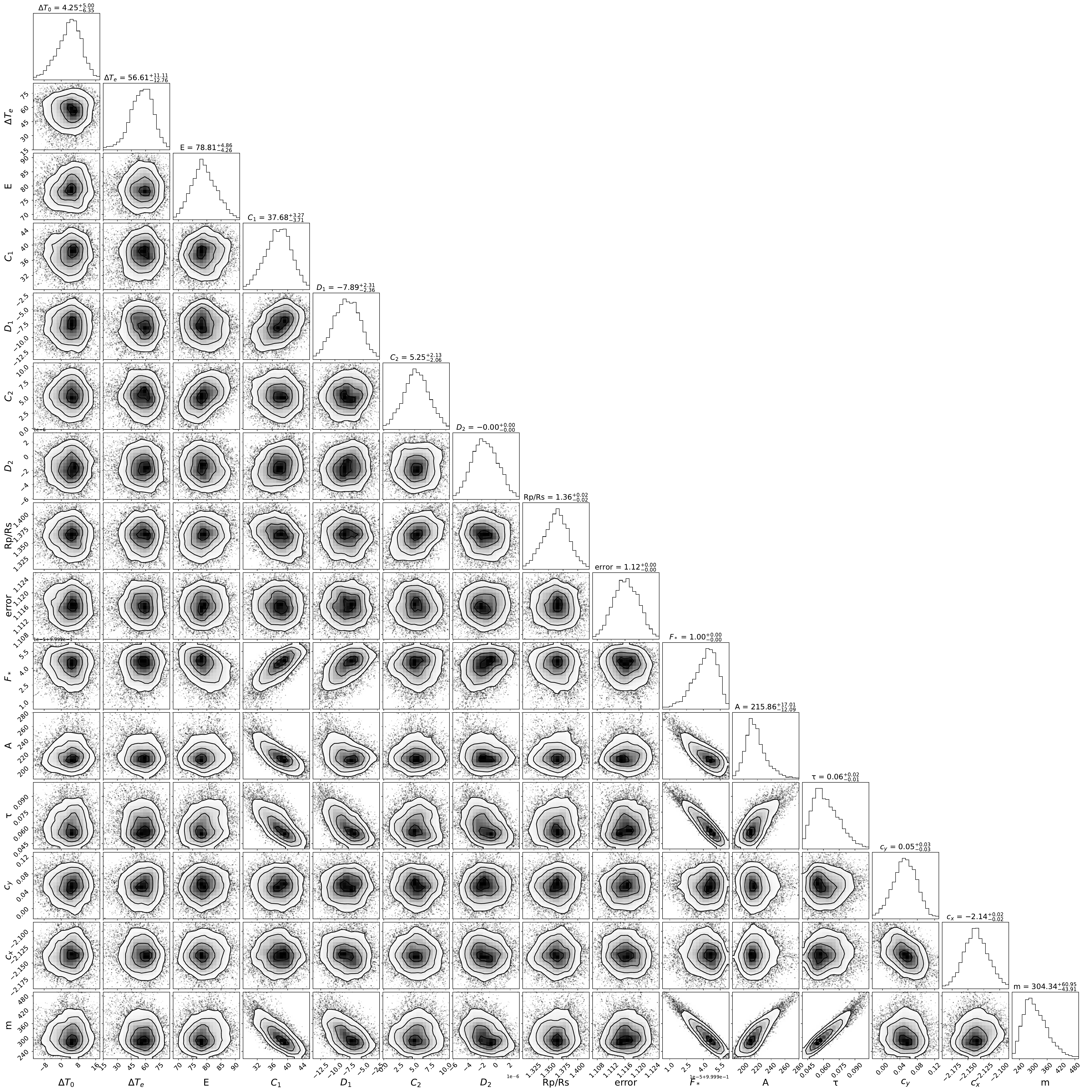}
    \caption{2D posterior distribution from the white phase curve MCMC fit.  The parameters are: transit time offset (s), eclipse time offset (s), eclipse depth (ppm), phase curve parameters $C_1$, $D_1$, $C_2$, $D_2$ (ppm), $R_p/R_s$ (\%), error inflation factor, normalization factor $F_*$, ramp amplitude A (ppm), ramp timescale $\tau$ (d), position decorrelation parameters $c_y$ and $c_x$ (\%/pix), and slope (ppm/day).}
    \label{fig:white_corner}
\end{figure*}

Figure \ref{fig:white_corner} shows the 2D posterior distributions (``corner plots'') of the parameters from the white phase curve fit.  Figure \ref{fig:amplitude_offset_corner} shows a corner plot of the amplitude and phase offset, which are derived from the parameters in the fit.

\section{The spectroscopic phase curve}
\label{sec:appendix_spectroscopic_phase_curve}

\begin{figure*}[ht]
    \subfigure {\includegraphics[width=0.5\textwidth]{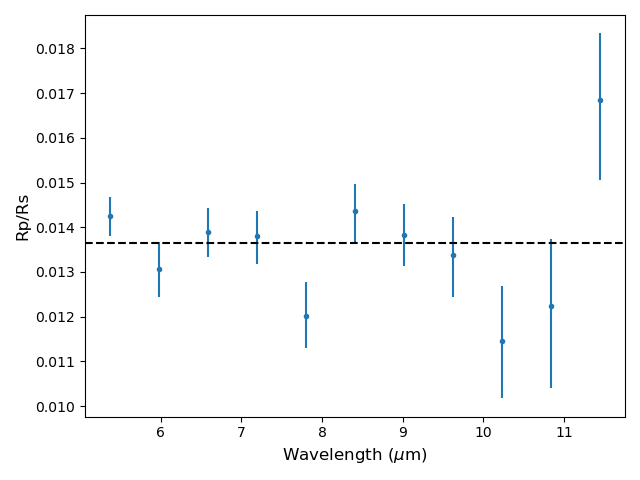}}\subfigure {\includegraphics[width=0.5\textwidth]{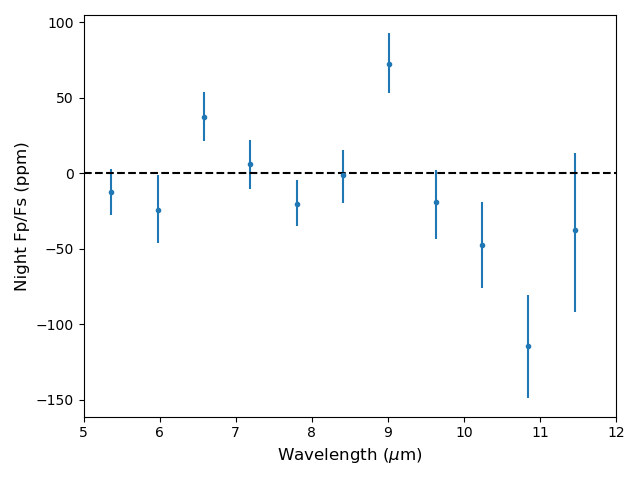}}
    \subfigure {\includegraphics[width=0.5\textwidth]{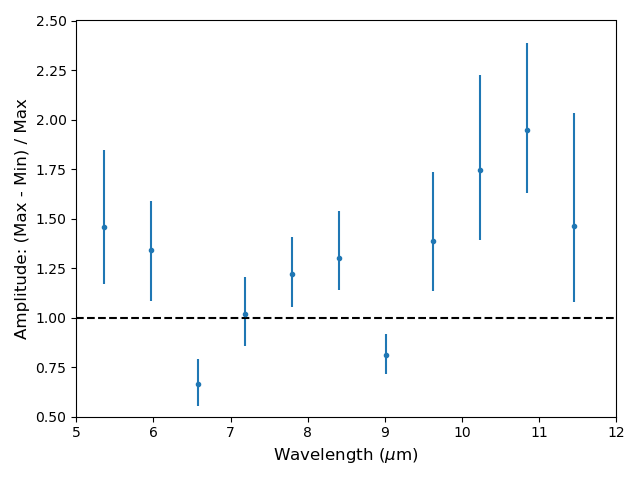}}\subfigure {\includegraphics[width=0.5\textwidth]{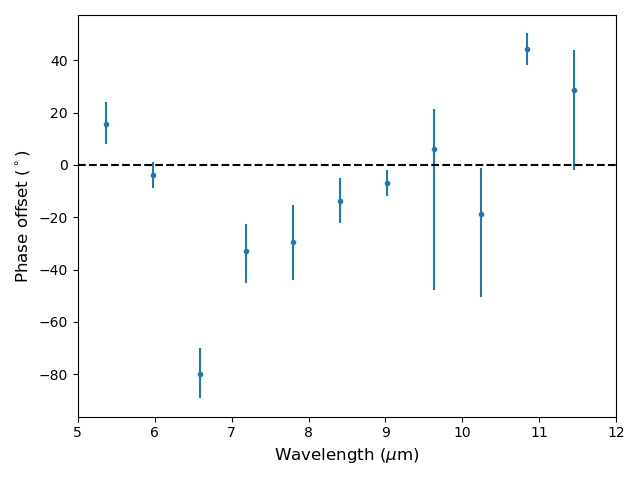}}
    \caption{Inferred parameters as a function of wavelength: transit depth, nightside emission, phase curve amplitude, and phase offset (negative means eastward).  On all plots, the expected value for an airless world with no heat recirculation is indicated by the dashed horizontal line.  On the transit spectrum, the line indicates the transit depth inferred from the white light curve.}
    \label{fig:spec_phase_curve}
\end{figure*}

In this main text of this paper, we focus on analyzing the white phase curve and the dayside emission spectrum.  We believe these are the most reliable data products. The emission spectrum is relatively robust because it is derived from two eclipses spanning a small fraction of the phase curve, reducing the impact of long-timescale instrumental or stellar variability.  The white phase curve is more robust than the spectroscopic phase curves because the MIRI/LRS transit spectrum of L168-9b, taken during commissioning, shows that bigger wavelength bins are better \citep{bell_2024}: the scatter in the transmission spectrum decreases faster with increased wavelength binning than 1/$\sqrt{N_{\rm bin}}$.  The cause of this phenomenon is unknown, but could be related to the odd-even effect or a 390 Hz electronic noise observed in several MIRI subarrays \citep{bell_2024}.

Figure \ref{fig:spec_phase_curve} shows the parameters derived from the spectroscopic phase curve: the transit spectrum, nightside emission spectrum, phase curve amplitudes, and phase offsets.  The transit spectrum is the most robust of the four, as it is effectively based on only a small portion of the phase curve.  The spectrum is marginally consistent with being flat, at a significance level of 0.05 ($\chi^2=20.7$, 12 dof).  The nightside emission spectrum is mostly consistent with zero, except at 9 \um, where the unphysically high eclipse depth results in an unphysically high value for the nightside emission, and at 10.8 \um, where the spectrum is highly negative (and therefore unphysical).  Most of the wavelength-dependent amplitudes are consistent with 1 at 2$\sigma$, but prefer values greater than 1--which would imply a negative planet flux at some phases, a physical impossibility.  Likewise, while most of the phase offsets are consistent with 0 at 2$\sigma$, some have very implausible values, such as 80$^\circ$ East at 6.6 \um or 40$^\circ$ West at 10.8 \um.  Unlike for the nightside flux or phase amplitude, we cannot conclusively prove that any phase offset is physically impossible.  We include these inferred spectroscopic phase curve parameters in the Appendix in the hopes that future scientists will be able to explain them with improved instrumental or astrophysical models.

\section{Alternate analysis: ignoring the last group}
\begin{figure*}[ht]
    \subfigure {\includegraphics[width=0.5\textwidth]{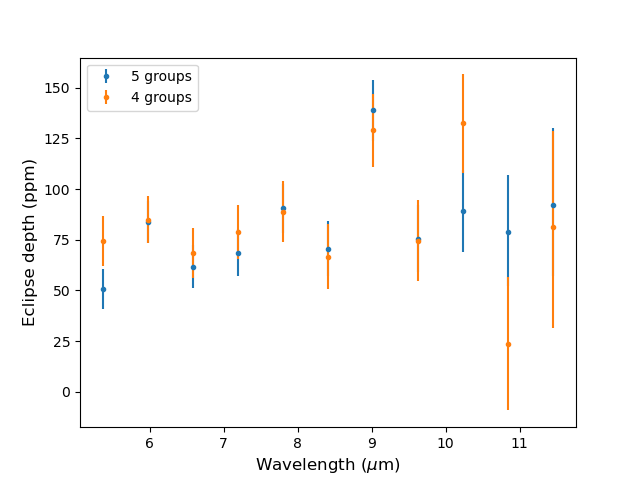}}
    \subfigure {\includegraphics[width=0.5\textwidth]{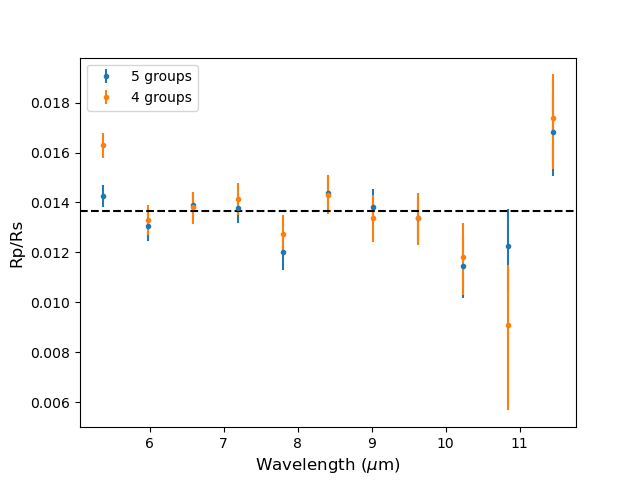}}
    \caption{Emission and transmission spectra obtained by using all five groups (up-the-ramp reads), compared to those obtained by using only the first four groups.  The former is the analysis we adopt.}
    \label{fig:ignore_last_grp_comp}
\end{figure*}

Due to the brightness of GJ 367, we were forced to use only 5 groups (up-the-ramp reads) per integration to avoid saturation.  The first five groups of MIRI integrations are known to be plagued by the reset switch charge decay, and all bright pixels are affected by the brighter-fatter effect \citep{wright_2023,morrison_2023}.  Additionally, the last group is known to be anomalous, with a downward offset proportional to the signal in the previous group \citep{morrison_2023}.  All of these effects introduce uncorrected non-linearity, which we attempt to compensate for using our two-step up-the-ramp fitting process (described in \citealt{kempton_2023} and Section \ref{sec:appendix_pipeline}).

To explore the sensitivity of our analysis to these various effects, we ran an alternate reduction which discards the last group.  Figure \ref{fig:ignore_last_grp_comp} shows the emission and transmission spectrum we obtain, compared to the fiducial analysis with all five groups.  The two are in agreement at most, but not all, wavelengths.  The four-group transmission spectrum is less flat--in particular, it has an anomalously high transit depth at 5.4 \um that exceeds the white light transit depth by 2.8 scale heights (assuming, implausibly, a H/He atmosphere), or $\sim$20 scale heights (assuming a high mean molecular weight atmosphere).  In the emission spectrum, the two analyses agree to within 1$\sigma$ at all but the shortest wavelength.

Our fit to the four-group white light curve shows an eclipse depth of $87 \pm 5$ ppm, an amplitude of $1.14 \pm 0.11$, and a phase offset of $-7.7_{-4.7}^{+3.9}$ degrees.  An amplitude greater than 1 is physically impossible, but the amplitude is consistent with 1 to 1.3$\sigma$.  The implied dayside temperature of $1824 \pm 100$ K is also higher than the maximum possible dayside temperature of $1720 \pm 40$ K, although the difference is not significant.  The phase offset is slightly smaller than in our fiducial analysis, but still 2$\sigma$ from 0.

In summary, the four-group analysis, like our fiducial analysis, is consistent with a dark airless world.  However, it results in a less physically plausible transmission spectrum, emission spectrum, and white phase curve.


\section{Is the 9\um spike a silicate feature?}
\label{sec:possible_silicate_feature}

\begin{figure*}[ht]
    \subfigure {\includegraphics[width=0.5\textwidth]{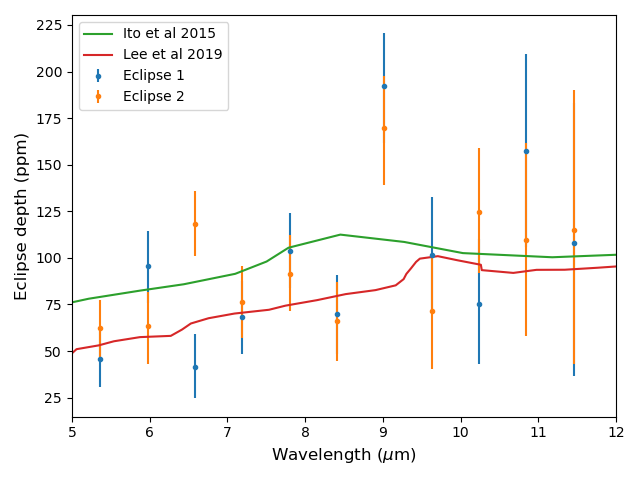}}\subfigure {\includegraphics[width=0.5\textwidth]{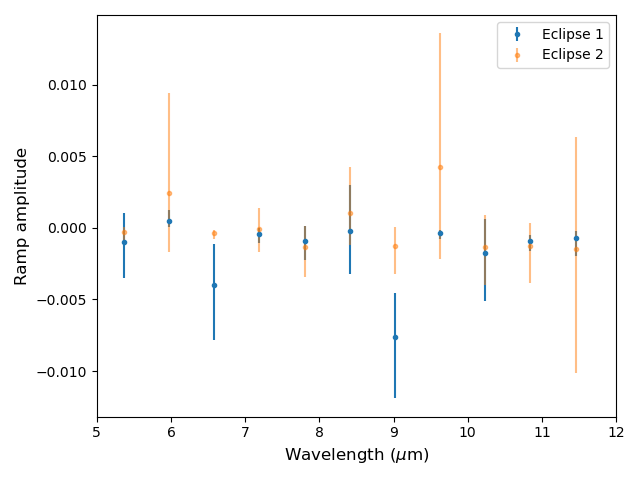}}
    \caption{Left: An alternate fit to the eclipse spectra, where the systematics model includes the ramp.  Note that the two eclipses both have a spike at 9 \um, and that the two spikes are consistent.  Comparing to scaled models that have a silicate emission feature at 9 \um \citep{ito_2015,lee_2019}, we see that the spike is much higher and sharper than silicate emission features found in these models. Right: in eclipse 1, 9 \um is unusual in that it has the highest ramp of all wavelengths.  In eclipse 2, there is nothing unusual about the ramp at 9 \um.}
    \label{fig:silicate_feature}
\end{figure*}

In this paper, we have been assuming that the 9\um spike in the emission spectrum is due to systematics.  However, one of our reductions casts severe doubt on this interpretation.  

The fiducial emission spectrum is made by cutting out the two eclipses from the data and fitting each separately, with a linear slope but no exponential ramp as the systematics model.  Following this methodology, eclipse 2 has an anomalously high depth at 9 um, but eclipse 1 is only somewhat high at that wavelength (see Figure \ref{fig:eclipses_comparisons}).

However, eclipse 1 is early enough in the observations that an exponential ramp might still exist.  When we include an exponential ramp in the systematics model, both eclipses are anomalously high at 9 \um, and the two eclipse depths are consistent (Figure \ref{fig:silicate_feature}).  We checked to see if there is anything unusual about the ramp amplitude or timescale at \um.  For eclipse 2, the answer is no, and the emission spectrum is almost identical whether we include the ramp or not.  In eclipse 1, however, the answer is yes: the ramp amplitude is substantially bigger at 9 um than at any other wavelength.  Adding the ramp to the fit does not significantly change the eclipse depth at any wavelength except 9 \um.

The facts above seem to rule out both instrumental \textit{and} astrophysical explanations.  If the spike is astrophysical, it would be a striking coincidence that this wavelength bin happens to have the biggest ramp in eclipse 1.  If the spike is instrumental, it would be an even more striking coincidence that two different systematic effects (the ramp in eclipse 1, and an unknown effect in eclipse 2) happened to cause the same anomalously high eclipse depth at the same wavelength in two different eclipses.

If the spike is indeed planetary, it may be due to silicates.  9 \um is the location of a Si-O stretching mode, which causes an absorption feature in quartz, enstatite, and other silicates (e.g. \citealt{jun_2003,kitzmann_2018,lee_2019}) as well as in gaseous SiO \citep{ito_2015,zilinskas_2023}.  Seeing a silicate feature in emission is not unexpected--a temperature inversion combined with silicate clouds or gaseous SiO would create such a feature.  A strong temperature inversion is not unexpected, as silicates are strong optical absorbers \citep{zilinskas_2023}.  It is also possible that volcanic eruptions eject hot gas, creating a tenuous atmosphere hotter than the surface \citep{heng_2023}.  However, while the existence of an emission feature at 9 \um is easy to explain, its height and narrowness are not: the emission features in \cite{ito_2015} and \cite{lee_2019} are much lower and broader compared to our observations (Figure \ref{fig:eclipses_comparisons}).  Since different silicates have different Si-O absorption bands, it is possible that some silicates (e.g. TiO$_2$; \citealt{kitzmann_2018}) have a sufficiently narrow absorption feature.  Even so, the exceptionally high brightness temperature of the spike--2800 K--is difficult to attain, and it is still more difficult for silicates to remain solid at these temperatures.  We encourage the community to explore the physical feasibility of such a strong inversion with more sophisticated models.

\bibliographystyle{aasjournal} \bibliography{main}
\end{document}